# Evaluation of Energy-Efficient Heuristics for ACO-based Routing in Mobile Wireless Sensor Networks

Arliones Hoeller Jr.
Telecommunications Department
Federal Institute of Technology of Santa Catarina
Florianópolis, Brazil

Antônio Augusto Fröhlich
Software/Hardware Integration Lab
Federal University of Santa Catarina
Florianópolis, Brazil

*Abstract*—Mobile ad-hoc networks demand routing algorithms able to adapt to network topologies subject to constant change. Moreover, with the advent of the Internet-of-Things (IoT), network nodes tend not only to show increased mobility, but also impose further restrictions to these mobile network systems. These restrictions include non-functional requirements of low-power consumption, small-size, and low-cost. ADHOP (Ant-based Dynamic Hop Optimization Protocol) is a routing algorithm based on ant-colony optimizations that target such small-size and low-cost platforms, consuming little amounts of memory and processing power. This paper elaborates on ADHOP to investigate the use of energy-related heuristics to guide routing decisions. The goal is to minimize network usage without jeopardizing network operation. We replace the original ADHOP heuristic of network latency by two energy-related heuristics based on the battery charge and the estimated node lifetime. Simulations compare the energy-aware versions of ADHOP to its original version, to AODV and to AOER. Results show that the proposed approaches are able to balance network load among nodes, resulting in a lower number of failures due to battery depletion. The energy-aware versions of ADHOP also deliver more packets than their counter-parts in the simulated scenario, delivering 2x more packets than the original ADHOP, and, respectively, 5x and 9x more packets than AOER and AODV.

## I. INTRODUCTION

The mobile ad-hoc networks in which packets need to perform multiple hops to reach their destination, as happens in most wireless sensor networks with mobile nodes [11], demand routing algorithms able to deal with the constant change in connectivity and limited resources, energy in particular. A successful mobile ad-hoc network must adapt to changes in connectivity, implementing mechanisms to detect failing links and new routes [4]. Besides node movement, battery depletion also is a major cause of link failure in these networks [7], making it important to handle energy consumption properly to avoid further degradation of connectivity.

AODV (Ad-hoc On-demand Distance Vector) is the most used routing protocol for mobile ad-hoc networks. Because it is a distance-vector routing protocol, all nodes in an AODV-based network must notify the changes in topology to their neighbors periodically. In mobile ad-hoc networks, especially in mobile wireless sensor networks, the control overhead generated by distance-vector protocols contrast with severe requirements of low-power operation. Indeed, AODV does not consider energy-related issues when making its routing decisions. To overcome this issue, several protocols have proposed the use of metaheuristic approaches in routing algorithms for mobile ad-hoc networks, and ACO (Ant Colony Optimization) [3] has proved to be a reliable tool in the design of these algorithms. For instance, AACS [10] uses ACO to build the minimum Steiner tree of the directed diffusion [8] routing mechanism, and AOER [15] uses ACO to minimize the number of hops in the routes. Although showing better results than AODV, these algorithms either do not integrate energy-related information to their heuristics or do it in a shallow way.

This paper presents EA-ADHOP (Energy-Aware ADHOP), a routing algorithm that uses specific energy-related information as heuristics for a routing mechanism based on the (ACO) method. The heuristics are included in ADHOP (Ant-based Dynamic Hop Optimization Protocol) [12], a routing protocol for mobile ad-hoc networks. ADHOP uses in its design different heuristics for different purposes, i.e. it can adapt itself to different types of networks to achieve a certain goal. The EA-ADHOP approach uses the adaptability of its algorithm to achieve energy efficiency without compromising delivery ratio. This algorithm can thereby work competitively to save nodes' energy or cooperatively to balance the use of energy across the network.

Simulations of the proposed algorithm were performed using two heuristic information to define routes: battery charge and expected lifetime. Simulations using OMNet++ show that, although the proposed versions of EA-ADHOP consume slightly more energy than the original ADHOP, they feature significantly lower standard deviation of energy consumption, suggesting a better distribution of energy usage among nodes.

EA-ADHOP versions also deliver 2x more packets than the original ADHOP, and, respectively, 6x and 10x more packets than AOER and AODV in dense networks with high degree of mobility.

The paper is organized as follows. Section 2 presents the related work. Section 3 presents the ADHOP approach for routing packets in WSNs. Section 4 shows the extensions to ADHOP to make it energy-aware. Section 5 shows simulation results to validate the proposed energy-aware ADHOP. Section 6 makes some final remarks.

## II. RELATED WORK

Routing algorithms for wireless sensor networks employ techniques to reduce energy consumption. Although featuring low-power implementations, not all of these algorithms use information about nodes' energy status to guide their routing decisions. In fact, those algorithms using such information to make routing decisions are often able to consume considerably less energy than those not using. That reduced energy consumption, however, comes at the expense of degraded network quality/performance. The routing algorithms described in this section are examples of those that, as ADHOP, use energy-related information explicitly in the decision-making process and propose ways to handle the energy-performance tradeoffs.

AACS (Adaptive Ant Colony System) [10] is a data-centric routing protocol that uses the directed diffusion [8] approach to route data packets. The directed diffusion method builds a MST (Minimum Steiner Tree) using a cost metric. AACS assumes that the amount of energy spent in a transmission is proportional to the number of hops needed to send a data packet from its source to its destination. Therefore, the energy-related heuristic information used to build the energy-efficient MST is the number of hops. AACS forwards data packets through the MST aggregating data in the tree's joining points, thus reinforcing the use of the preferred paths and reducing network load. When compared to traditional directed diffusion, AACS modifies it by using the ACO method to build the MST. Other methods build the MST for the directed diffusion scheme using DDSP (Destination-Driven Shortest Path) [2], DDMC (Destination-Driven Multicast) [14], and SBPT (Shortest Best Path Tree) [6]. These implementations, however, are less suitable to the scenario of mobile wireless sensor networks. The authors report that AACS reduces traffic thanks to aggregation and increased number of successful deliveries, thus saving energy and prolonging network lifetime. However, in mobile networks where topologies change too often, failing data delivery due to outdated MSTs and the overhead for keeping the MST updated overcomes the gains in energy consumption, jeopardizing both network lifetime and quality.

AODV (Ad-hoc On-demand Distance Vector) [13] is a reactive routing protocol based on distance vectors. It is widely used in mobile ad-hoc networks. AODV establishes routes by exchanging control messages to request routes (RREQ), reply with a route (RREP), or report a routing error (RERR). The transmission of RREQ via flooding and the use of HELLO messages to check route status generate significant control overhead. A simplified version of this protocol, called AODVjr, has shown to be more suitable for wireless sensor network scenarios, reducing network traffic. AODVjr is a trimmed down AODV specification that removes all but the essential elements of AODV [1]. The removed features include sequence numbers, hop-count, HELLO and RERR messages. Also, the number of RREP reduces drastically once AODVjr only allows the final destination to reply RREQ messages, meaning that RREPs are not sent from intermediate nodes that would know a route to the destination. A timeout mechanism that disables routes that remain inactive for long periods replaces the HELLO messages.

AOER (Ant-based On-demand Energy Route) [15] is a routing algorithm for mesh networks based on the IEEE 802.15.4 standard. Its implementation focuses at reducing intra-node overhead using simpler data structures and algorithms when compared to other ACO-based solutions, resulting in reduced use of memory, CPU, and, consequently, power. The ACO-based routing mechanism in AOER uses energy-related information to compute the ACO metaheuristic. This heuristic is a function of residual energy in the nodes' neighborhood, in the route, and on the whole network. AOER also has a safety mechanism to prevent battery depletion on over-used routes. This mechanism checks, on each node, the impact of each route the node serves on energy consumption. If the energy consumed by a node increases too much due to the establishment of a new route, AOER disables the route and starts a new route discovery procedure. AOER case-studies show significant performance enhancement compared to AODVjr, featuring less energy consumption and lower variance of residual energy among nodes, suggesting a better load distribution.

## III. ADHOP

ADHOP (Ant-based Dynamic Hop Optimization Protocol) [12] is a routing protocol for mobile ad-hoc networks that uses ACO (Ant Colony Optimization) [3] to route packets. Figure 1 shows a general view of this mechanism. In ADHOP, ants are piggybacked to data packets. These ants record the route they take and collect metadata about the links and nodes they pass. After reaching a destination, the ants return to the source through the same route they took, depositing pheromone in the links on their way back. As time goes by, pheromone evaporates.

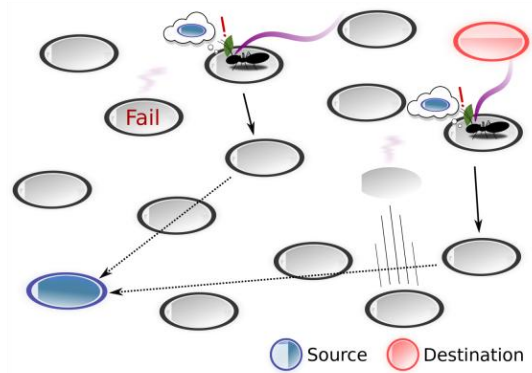

Figure 1. ACO-based route selection in ADHOP.

Figure 2. ADHOP Routing Table Structure.

ADHOP uses metadata brought by the ants to define the amounts of pheromone deposited in a link and the pheromone evaporation rate. ADHOP selects routes to forward packets based on the pheromone levels on the route's first link – the preferred routes are those with higher levels of pheromone. This mechanism allows ADHOP to adapt to changing topologies: since ants will not return unless they successfully reach a destination, pheromone levels on broken routes will gradually reduce, forcing ADHOP to explore alternate routes.

The equation used to update the amount of pheromone in a route is [3]:

$$\tau_{A,B} = (1 - \phi) \cdot \tau_{A,B} + \phi \cdot \tau_0 \quad (1)$$

where $\phi \in [0, 1]$ is the pheromone decay coefficient, $\tau_0$ is the initial amount of pheromone in the route, and $\tau_{A,B}$ is the current amount of pheromone in the route. Pheromone evaporation occurs periodically in all routes according to the equation [3]:

$$\tau_{A,B} = (1 - \rho) \cdot \tau_{A,B} \quad (2)$$

where $\rho \in [0, 1]$ is the evaporation rate.

ADHOP labels each entry of the routing table with a pair of node identifiers. For instance, the entry (A, B) labels a route to node A using the link with the neighbor node B. The structure of the ADHOP Routing Table is similar to the structures used by HOPNET Intra Routing Table [16] and AOER Inverted Routing Table [15]. In ADHOP, however, operations over the routing table are simpler and faster than in HOPNET and AOER. This happens because ADHOP does not store the entire routes, thus reducing in-node memory usage and the size of the data structures. Additionally, ADHOP uses a hash table where destination addresses are the keys.

Figure 2 shows an instance of ADHOP routing table for a hypothetic node i. Each bucket in the hash table holds a list of entries sorted by the quantity of pheromone on the link. Each entry in the lists keeps the neighbor node (prefixed by "Nei.") to which a packet should be forwarded to reach its destination (prefixed by "Dst."). The algorithm will always choose the first entry matching the final destination of the packet, i.e. the route with more pheromone and, therefore, with the higher probability of performing a successful delivery. The neighbor can be the targeted destination as happens in the figure with destinations 00 and 06, or a hop in the direction of the destination.

ADHOP uses two types of ants to perform operations over the routing tables: Forward Transport Ant (FTA) and Exploratory Transport Ant (ETA). FTAs are used to forward data through known routes, while ETAs are broadcasted by nodes that do not know routes to the ants' destinations, serving as a reactive route exploration mechanism. Both ants share the data structure shown in Figure 3. The data structure includes traditional fields such as the Source and Destination addresses. The Previous field stores the address of the previous node. ADHOP uses the SequenceNO field for sequence control. The Type field indicates the ant category, and the Hops field indicates the number of hops the ant has done. The field Heuristic Inf. stores the necessary heuristic information to support the pheromone evaporation and deposit functions.

The UML sequence diagram in Figure 4 shows the data transmission procedure. Both FTA and ETA deliver data while depositing pheromone on the routes they use. This ensures that sudden changes in the network topology do not interfere with data transportation. Ants unwittingly notice the changes in topology as the amount of pheromone in a link varies.

Figure 5 shows a UML sequence diagram depicting the reception of a FTA. Upon the reception of a FTA, if the packet has reached its destination, ADHOP delivers the data to the upper layer. If the destination has not been reached yet, ADHOP looks up for the next hop in ADHOP-RT and forwards the FTA. It forwards a new ETA if there is no known route to the destination.

ETAs are responsible for discovering routes to unknown nodes. Each node retransmits the ETAs to all of its neighbors until the ant reaches either the destination or a node that has a route to the destination. At the destination, the ETA delivers the data packet and returns to the source node. Figure 6 shows the sequence diagram for the reception of an ETA. On the way back, the ant reinforces the pheromone trail.

It is possible to extend the routing mechanism of ADHOP using the Heuristic Inf. field of the protocol header (Figure 3) to transmit relevant information. The information exchanged through the Heuristic Inf. field can be used in modified versions of the equations for deposit and evaporation of pheromone (Equations 1 and 2). For instance, the battery charge of the neighbor nodes can affect the deposit function so that links to nodes with more energy have more pheromone than links to nodes with less energy. This would balance energy utilization across the network and help to extend the lifetime of the network as a whole.

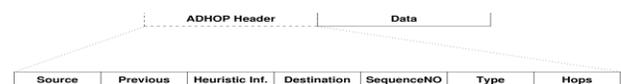

Figure 3. ADHOP Ant Structure.

These modifications are better discussed in Section 4.

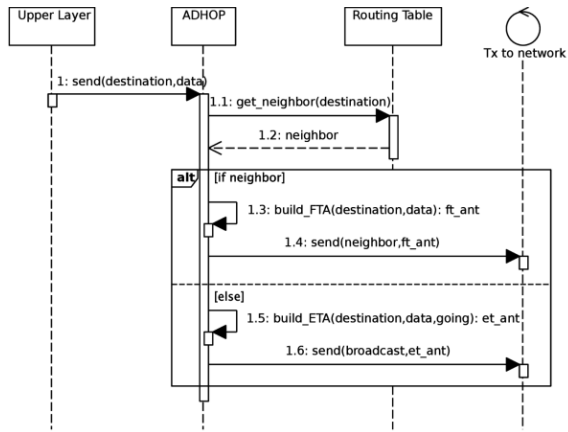

Figure 4. ADHOP - Data Transmission.

## IV. ENERGY-AWARE ADHOP

As stated in section 3, ADHOP uses the ACO metaheuristic optimization mechanism, meaning that different heuristics can be used to adapt the system according to different goals. The EA-ADHOP approach aims at using this adaptability to achieve energy efficiency without sacrificing delivery ratio. This algorithm can thereby work competitively to save nodes' energy or cooperatively to balance the use of energy across the network.

All energy information can be collected and shared involuntarily by ants as they traverse the network. In the Heuristic Inf. field (Figure 3), all ants attach some relevant information to the energy-aware routing, such as residual energy, expected battery lifetime, or consumed energy. The algorithm uses such information to calculate the pheromone deposit ratio and the evaporation ratio (Equations 1 and 2), as shown in Figure 7. For instance, EA-ADHOP can use the following cooperative heuristic: "Always transmit the data to the neighbor node with higher estimated lifetime". In this heuristic, the ants bring the estimated lifetime of the last neighbor node from which they came from. The φ value depends on the value assigned in the Heuristic Inf. field. Meanwhile, the algorithm decreases the evaporation ratio for the nodes with higher estimated lifetime by decreasing the ρ value.

In this paper, EA-ADHOP was modeled to use two distinct energy-aware heuristics as metrics to guide the routing. The first metric considered is the residual energy. In this configuration, the Heuristic Inf. field of ants receives a value representing the current state-of-charge of the node's battery. The range of this value is (0, 1], representing the battery charge ranging from 0% to 100%. The battery charge metric is viable in WSNs where the energy source is symmetric, i.e. in WSNs where every battery in every node has the same size. For instance, suppose two nodes in a network, being one powered by a battery of 1000 mAh and another one with a battery of 200 mAh. If the first is 30% charged, it has 300 mAh remaining, and if the second is 50% it has 100 mAh remaining. In this scenario, EA-ADHOP would deposit more pheromone in the link to the node with less energy, once the relative charge on that node would be larger. This would render undesired behavior of the pheromone control mechanism.

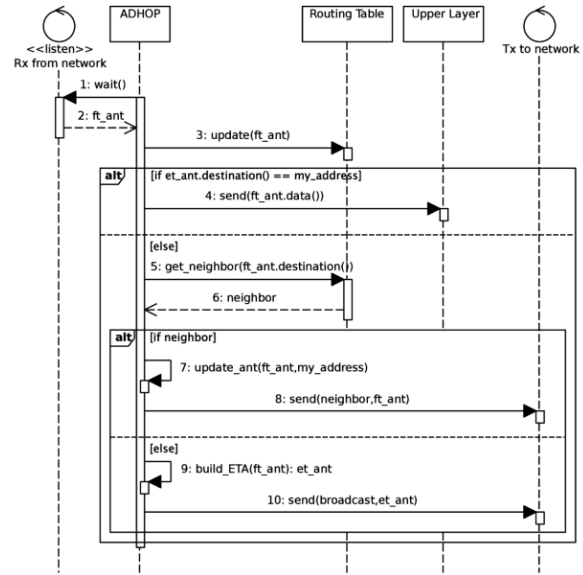

Figure 5. ADHOP - Reception of a Forward Transport Ant (FTA).

To cope with heterogeneous network nodes, the use of estimates of battery lifetime instead of residual energy is proposed. The EA-ADHOP using the battery lifetime heuristic assumes that there is a targeted operation lifetime for the network. Thus, the Heuristic Inf. field of ants receive a value representing the proximity of the estimated lifetime of a node to its targeted lifetime. The range of this value also is (0, 1], representing the percentage of the estimated lifetime of a node compared to the targeted lifetime of the system. The approach for defining the estimated battery lifetime is in Equations 3, 4, and 5. Firstly, the lifetime estimator defines the rate at which the system consumes energy ($D_i$). The discharge rate is the rate of energy consumed over time. Afterwards, the division of the current battery charge by the computed discharge rate will give the estimated lifetime ($\hat{L}$). Finally, there is a correction of the estimated lifetime to the (0, 1] range performed by dividing $\hat{L}$ by the time remaining to reach the targeted lifetime ($L_T$). Whenever the estimated lifetime is larger than the targeted lifetime, the heuristic metric value ($H_L$) assumes the value of one.

$$D_i = \frac{E_{batt}^0 - E_{batt}^i}{t_i} \quad (3)$$

$$\hat{L} = \frac{B_i}{D_i} \quad (4)$$

$$H_L = min\left(1, \frac{\hat{L}}{(L_T - t_i)}\right) \quad (5)$$

ADHOP must have access to energy-related information to build energy-awareness. There are three facets of energy that must be taken into account in wireless sensor network systems: consumption, production, and storage. In this system, only consumption and storage are considered. This system uses a combination of measurements and event accounting to monitor energy [9].

Equations 6 through 8 show the employed energy consumption model. Depending on the behavior of a given component, or information available about its energy consumption, the designer chooses to monitor energy consumption based on either the time the device spends in each operating mode or the events the device generates. $E_{tm}^i(d, \Phi)$ defines the energy consumed by a single device (d) over time as a function of current drain (I) and time (t) spent in an operating mode (m) with a given configuration ($\Phi$). $E_{ev}^i(d, \Phi)$ is the sum of the energy consumed by events that are relevant in terms of energy consumption. During execution, the system accounts for these events ($\chi$). Each event has a known worst-case energy consumption ($E_e(\Phi)$) subject to the system configuration ($\Phi$). In the case of a device consuming energy in both ways, both energy consumption profiles can be applied. Finally, $E_{tot}^i(\Phi)$ is the sum of the estimated energy consumption of all system devices ($\Delta$), during the ith iteration.

$$E_{tm}^i(d, \Phi) = (t_{end} - t_{begin}) \times I(d, m, \Phi) \quad (6)$$

$$E_{ev}^i(d, \Phi) = \sum_{e=0}^{|\chi|} (E_e(\Phi) \times \chi_e) \quad (7)$$

$$E_{tot}^i(\Phi) = \sum_{d=0}^{|\Delta|} (E_{tm}^i(d, \Phi) + E_{ev}^i(d, \Phi)) \quad (8)$$

$$E_{batt}^i = E_{batt}^{(i-1)} - E_{tot}^i(\Phi) \quad (9)$$

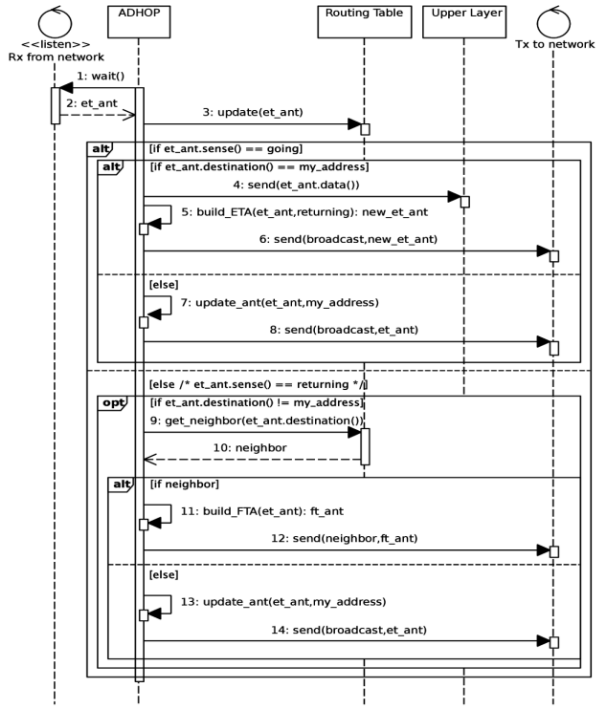

Figure 6. ADHOP - Reception of an Exploratory Transport Ant (ETA).

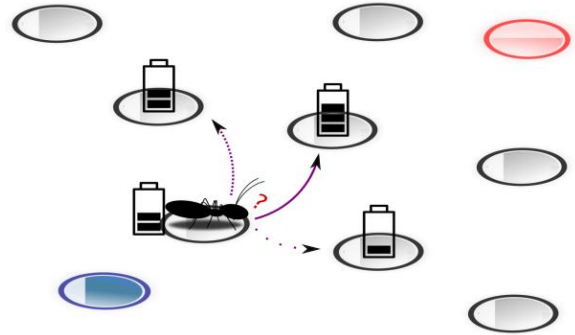

Figure 7. EA-ADHOP - Choosing the node with higher battery charge.

$E_{tm}^i(d, \Phi)$ is updated either on every operating mode change or periodically, when $E_{ev}^i(d, \Phi)$ and $E_{tot}^i(\Phi)$ are also updated. The period of the iterations (i) vary from application to application and has already been subject of previous studies [9]. The amount of energy available to the system in a given moment ($E_{batt}^i$) can be estimated as shown in Equation 9. Given a previously known charge of the battery ($E_{batt}^{(i-1)}$), current battery charge comes from the subtraction of the amount of energy consumed in a period ($E_{tot}^i(\Phi)$).

V. EVALUATION AND RESULTS

The experiments use the OMNeT++ – an extensible, modular, component-based C++ framework for building network simulations. The simulation parameters follow the characteristics of a specific wireless sensing module: the EPOSMote [5]. The EPOSMote is a modular platform for building wireless sensor network applications. It features a low-power System-on-a-Chip (SoC) that includes a RF transceiver compatible with the IEEE 802.15.4 standard and an integrated 32-bit ARM7 processor. Table 1 presents relevant parameters of the power profile of EPOSMote.

TABLE I. CURRENT DRAIN OF CPU AND RADIO IN EPOSMOTE.

| Device/Mode | Current Drain |
| --- | --- |
| CPU Active Current | 3.3 mA |
| CPU Sleep Current | 60 µA |
| CPU Hibernate Current | 0.9 µA |
| Radio Tx Current | 28.3 mA |
| Radio Rx Current | 21.3 mA |
| Radio Sleep Current | 0.1 µA |

Table 2 shows the OMNeT++ simulation parameters. In these experiments, each simulation scenario ran for 900

seconds in an environment of high mobility that is conducive to high data loss. The simulation places nodes randomly in a squared area of 1.44 km2 (edges of twelve hundred meters), and each node moves at a maximum speed of five meters per second, according to OMNeT++'s Mass Mobility profile. Twenty mobile source nodes generate data traffic to other twenty mobile sink nodes. The experiment explores the behavior of the routing algorithms when varying the number of routing nodes (ranging from twenty to two hundred). The experiments evaluate the energy-aware versions of ADHOP using the two heuristic metrics: battery charge (EAADHOP-B) and estimated lifetime (EA-ADHOP-L).

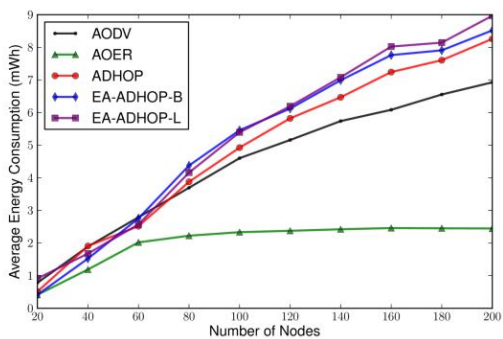

Figure 8. Average energy consumption.

TABLE II.　OMNET++ CONFIGURATION.

| Parameter | Value |
|---|---|
| Simulation Time | 900 seconds |
| Number of Nodes | 20 ~ 200 |
| Area | 1200m X 1200m |
| Mobility Model | Mass Mobility |
| Application Message Length | 32 bytes |
| Application Message Frequency | 0.25 Hz (every 4s) |
| UDP Header Length | 8 bytes |
| IP Header Length | 20 bytes |
| Netmask | 255.255.0.0 |
| ADHOP Header Length | 20 bytes |
| IEEE 802.15.4 ACK | True |
| IEEE 802.15.4 Header Length | 22 bytes |
| IEEE 802.15.4 Max Frame Size | 102 bytes |
| PHY Transmitter Power | 1 mW |
| PHY Sensitivity | −85 dBm |
| PHY Thermal Noise | −110 dBm |
| Channel Carrier Frequency | 2.4 GHz |
| Battery Voltage | 3 V |

The results show how the EA-ADHOP compares to three other routing algorithms: ADHOP, AOER, and AODV. In the simulations, the battery capacity parameter was reduced to match the simulation length. This is needed because, if the battery size is too large, there will not be enough time for the simulations to show relevant variations in the heuristics. The parameters related to the energy source used in the simulation are in Table 3.

Figure 8 shows that the average energy consumption of both versions of EA-ADHOP are similar to compared approaches. However, the reduced standard deviations of the energy consumption shown in Figure 9 for the energy-aware versions of ADHOP suggest that there is a better distribution of traffic among nodes. Also, the graphics show that EA-ADHOP-L performs better than EA-ADHOP-B. This is because the response time of the battery lifetime metric is faster than that of the battery charge.

Figure 9 shows the standard deviation of the energy consumption of the nodes after 900 seconds of intense data transmission. On average, both EA-ADHOP approaches improve the balance of energy consumption when compared to the original ADHOP. AODV results in terms of standard deviation are similar to those of EA-ADHOP-B and EAADHOP-L

TABLE III.　OMNET++ BATTERY CONFIGURATION PARAMETERS.

| Battery Charge Heuristic Metric EA-ADHOP-B | |
|---|---|
| Battery Capacity | 3 mAh * 3V = 9 mWh |
| **Estimated Lifetime Heuristic Metric EA-ADHOP-L** | |
| Battery Capacity | 3 mAh * 3V = 9 mWh |
| Target Lifetime | 900 seconds |

Two extra observations must be done about the results in Figures 8 and 9. First, all algorithms show similar energy consumption and low standard deviation in the 20 nodes configuration. This happens because 20 nodes are too few to ensure connectivity among every node in the 1200x1200 meters grid. As a consequence, too many packets end up undelivered, as shown in Figure 10.

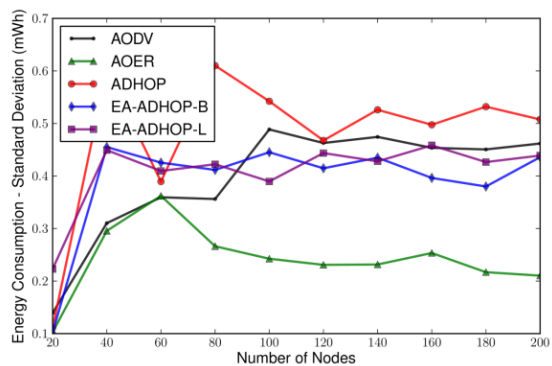

Figure 9. Standard deviation of energy consumption.

Second, AOER seems the best option in terms of average energy consumption and distribution of network activity (standard deviation of energy consumption). However, it is important to note that this result is a consequence of AOER's poor performance in terms of packet delivery (Figure 10). The AOER's reduced packet delivery ratio is due to its inability to keep up with the constant change in connectivity caused by mobility in this experiment.

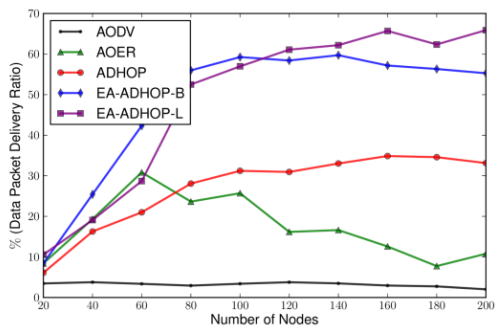

Figure 10. Delivery ratio of data packets.

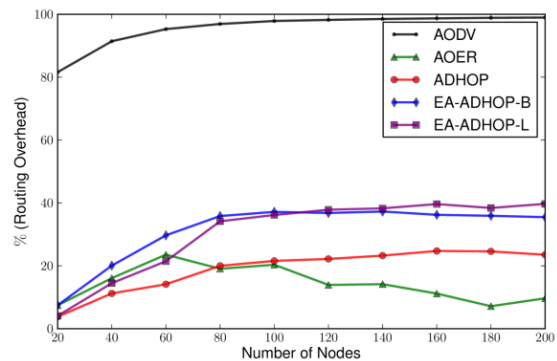

Figure 11. Routing overhead.

Figure 11 shows the routing overhead of the studied protocols. As expected, AODV show an extremely high routing overhead - around 99% for a larger number of nodes. This means that the AODV network transmits, roughly, 99 bytes of control data for every byte of useful data. As a consequence, the efficiency of the network is very low. This happens because AODV is a distance vector routing algorithm acting proactively to discover routes. In a network with mobile nodes moving too often, the exchange of routing information in AODV ends up flooding the network.

The proposed versions of EA-ADHOP show significantly higher routing overhead when compared to ADHOP and AOER. The reason for it is that, in EA-ADHOP, routes change not only because of mobility, but also because of battery discharge and consequent variations on expected battery lifetime. These changes in routes generates more exploratory ants (ETA), increasing overhead. However, this overhead does not seem to impact network performance. In fact, as Figure 10 shows, the EA-ADHOP-B and EAADHOP-L algorithms show the best packet delivery ratio. It is a consequence of the algorithms being able to distribute routes across the network, thus avoiding the overuse of specific routers. In the simulations using the EA-ADHOP algorithms, none of the nodes had their battery fully depleted. All other algorithms overused some routers, causing these nodes to fail due to battery depletion.

The results show that the EA-ADHOP algorithms not only homogenize the energy consumption on the network, but also enhance the data delivery ratio and, consequently, make the system more energy-efficient because it spends less energy per delivered packet.

## VI. CONCLUSION

This paper presented EA-ADHOP: an energy-efficient version of the ADHOP routing protocol for mobile wireless sensor networks. As in ADHOP, EA-ADHOP uses ant-colony optimization techniques to find routes in a mobile network. EA-ADHOP, however, uses energy-related information to achieve that, seeking to homogenize energy consumption among nodes. The energy-related heuristic metrics are used in the ant-colony optimizer of the algorithm to balance the functions defining the deposit and evaporation rates of pheromone in the network links. Two distinct metrics were explored: battery charge and estimated battery lifetime. The modeling and behavior of the routing algorithm is explained, as are the mechanisms to obtain the energy-related metrics.

Results show that EA-ADHOP enabled the system to sustain operation for a target lifetime, eliminating failures due to battery depletion and raising packet delivery rate. As a consequence, the resulting system is more energy-efficient as it spends less energy per delivered packet when compared to others.

On-going research is now exploring the use of ADHOP's ACO mechanism to adapt transmission power and reception sensitivity of sensor nodes. The same work is also exploring the impact of the adaptation of these operational parameters in routing efficiency. Future work involving will also evaluate the algorithm in different mobility scenarios.

### ACKNOWLEDGMENT

Authors would like to thank Mr. Alexandre Massayuki Okazaki, the author of the original ADHOP algorithm, for important discussion and technical support in the development of this work.

REFERENCES

[1] I. D. Chakeres and L. Klein-Berndt. Aodvjr, aodv simplified. SIGMOBILE Mob. Comput. Commun. Rev., 6(3):100–101, June 2002.

[2] E. Dijkstra. A note on two problems in connexion with graphs. Numerische Mathematik, 1(1):269–271, 1959.

[3] M. Dorigo, M. Birattari, and T. Stutzle. Ant colony optimization. Computational Intelligence Magazine, IEEE, 1(4):28–39, nov. 2006.

[4] C. Elliott and B. Heile. Self-organizing, self-healing wireless networks. In Personal Wireless Communications, 2000 IEEE International Conference on, pages 355–362, 2000.

[5] Federal University of Santa Catarina. The EPOSMote Project. Internet, oct 2012.

[6] H. Fujinoki and K. Christensen. The new shortest best path tree (sbpt) algorithm for dynamic multicast trees. In Local Computer Networks, 1999. LCN '99. Conference on, pages 204–211, 1999.

[7] N. Gupta and S. Das. Energy-aware on-demand routing for mobile ad hoc networks. In S. Das and S. Bhattacharya, editors, Distributed Computing, volume 2571 of Lecture Notes in Computer Science, pages 164–173. Springer Berlin Heidelberg, 2002.

[8] C. Intanagonwiwat, R. Govindan, and D. Estrin. Directed diffusion: a scalable and robust communication paradigm for sensor networks. In Proceedings of the 6th annual international conference on Mobile computing and networking, MobiCom '00, pages 56–67, New York, NY, USA, 2000. ACM.

[9] A. Hoeller Jr. and A. A. Fröhlich. On the Monitoring of System-Level Energy Consumption of Battery-Powered Embedded Systems. In 2011 IEEE International Conference on Systems, Man, and Cybernetics, pages 2608–2613, Anchorage, AK, USA, Oct. 2011.

[10] Z. Li and H. Shi. A Data-Aggregation Algorithm Based on Adaptive Ant Colony System in Wireless Sensor Networks. Congress on Image and Signal Processing, 4:449–453, 2008.

[11] L. Mottola and G. P. Picco. Programming wireless sensor networks: Fundamental concepts and state of the art. ACM Comput. Surv., 43(3):19:1–19:51, Apr. 2011.

[12] A. M. Okazaki and A. A. Frohlich. Ant-based dynamic hop optimization protocol: A routing algorithm for mobile wireless sensor networks. In GLOBECOM Workshops (GC Wkshps), 2011 IEEE, pages 1139 –1143, dec. 2011

[13] C. Perkins and E. Royer. Ad-hoc on-demand distance vector routing. In Proceedings. Second IEEE Workshop on Mobile Computing Systems and Applications, pages 90–100, feb 1999.

[14] A. Shaikh and K. Shin. Destination-driven routing for low-cost multicast. Selected Areas in Communications, IEEE Journal on, 15(3):373–381, 1997.

[15] B. Shuang, Z. Li, and J. Chen. An ant-based on-demand energy route protocol for ieee 802.15.4 mesh network. International Journal of Wireless Information Networks, 16:225–236, 2009.

[16] J. Wang, E. Osagie, P. Thulasiraman, and R. K. Thulasiram. HOPNET: A hybrid ant colony optimization routing algorithm for mobile ad hoc network. Ad Hoc Network, 7(4):690–705, 2009.